# A Special Non-linear Equation of Mathematical Physics – the Equation of the Curvilinear Electromagnetic Wave

Alexander G. Kyriakos

*Saint-Petersburg State Institute of Technology,*
*St. Petersburg, Russia*
*Present address: Athens, Greece, e-mail: agkyriak@yahoo.com*

**Abstract**
A special non-linear equation of curvilinear electromagnetic wave is presented. The particularity of this equation lies in the fact that in matrix form it is mathematically equivalent to the Dirac electron equation. It is shown that the solution of this equation is the motion of the plane-polarized electromagnetic wave on a circular trajectory. It is also shown, that such twirled wave can be considered as a massive charge particle with a spin of one half, similar to electron. The non-linear equation and its Lagrangian of these EM particles are founded.

## 1.0. Basic supposition

*Let us suppose that under some conditions a linear electromagnetic wave (photon) is able to twirl and move along the closed curvilinear trajectory.*

As is known according to Maxwell-Lorentz electromagnetic theory [1] those conditions appear when electromagnetic wave propagates in the very strong external electromagnetic field.

Below we translate this supposition on the mathematics language and show that in matrix form the non-linear equations of such curvilinear waves mathematically fully coincides with quantum equations.

## 2.0. Linear electromagnetic wave equation in the matrix form

Let us consider the plane electromagnetic wave moving, for example, on $y$ - axis (fig.1):

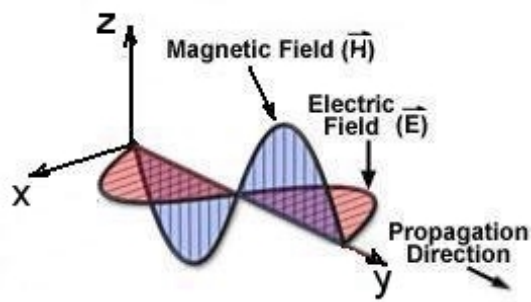

Fig. 1

where the electric and magnetic fields can be written in the complex form as:

$$\begin{cases} \vec{E} = \vec{E}_o e^{-i(\omega t \pm ky)}, \\ \vec{H} = \vec{H}_o e^{-i(\omega t \pm ky)}, \end{cases} \quad (2.1)$$

In the general case this electromagnetic wave of any direction has two polarizations and contains only four field vectors. For example in the case of y-direction we have:

$$\vec{F} = \{E_x, E_z, H_x, H_z\}, \quad (2.2)$$

and $E_y = H_y = 0$ for all transformations.

The electromagnetic wave equation has the following view [1]:

$$\left(\frac{\partial^2}{\partial t^2} - c^2 \vec{\nabla}^2\right)\vec{F} = 0, \quad (2.3)$$

where $\vec{F}$ is any of the above electromagnetic wave fields (2.2). In other words this equation represents four equations: one for each wave function of the electromagnetic field.

We can also write this equation in the following operator form:

$$\left(\hat{\varepsilon}^2 - c^2 \hat{\vec{p}}^2\right)\vec{F}(y) = 0, \quad (2.4)$$

where $\hat{\varepsilon} = i\hbar \dfrac{\partial}{\partial t}, \hat{\vec{p}} = -i\hbar \vec{\nabla}$ are the operators of the energy and momentum correspondingly.

Taking into account that $(\hat{\alpha}_o \hat{\varepsilon})^2 = \hat{\varepsilon}^2$, $(\hat{\vec{\alpha}} \hat{\vec{p}})^2 = \hat{\vec{p}}^2$, where $\hat{\alpha}_o = \begin{pmatrix} \hat{\sigma}_0 & 0 \\ 0 & \hat{\sigma}_0 \end{pmatrix}$; $\hat{\vec{\alpha}} = \begin{pmatrix} 0 & \hat{\vec{\sigma}} \\ \hat{\vec{\sigma}} & 0 \end{pmatrix}$; $\hat{\beta} \equiv \hat{\alpha}_4 = \begin{pmatrix} \hat{\sigma}_0 & 0 \\ 0 & -\hat{\sigma}_0 \end{pmatrix}$ are Dirac's matrices and $\hat{\sigma}_0, \hat{\vec{\sigma}}$ are Pauli matrices, the equation (2.4) can also be represented in the matrix form of the Klein-Gordon equation without mass:

$$\left[(\hat{\alpha}_o \hat{\varepsilon})^2 - c^2(\hat{\vec{\alpha}} \hat{\vec{p}})^2\right]\Psi = 0, \quad (2.5)$$

where $\Psi$ is some matrix, which consists four components of $\vec{F}(y)$.

Taking into account that in case of photon $\omega = \varepsilon/\hbar$ and $k = p/\hbar$, from (2.5), using (2.1), we obtain $\varepsilon = cp$, as for a photon. Therefore we can consider the $\Psi$ - wave function of the equation (2.5) both as EM wave and as a photon (such portion of the electromagnetic wave we can conditionally name an "EM-string").

Factorizing (2.5) and multiplying it from left on the Hermitian-conjugate function $\Psi^+$ we get:

$$\Psi^+\left(\hat{\alpha}_o \hat{\varepsilon} - c\hat{\vec{\alpha}} \hat{\vec{p}}\right)\left(\hat{\alpha}_o \hat{\varepsilon} + c\hat{\vec{\alpha}} \hat{\vec{p}}\right)\Psi = 0, \quad (2.6)$$

The equation (2.6) may be disintegrated on two Dirac equations without mass:

$$\Psi^+\left(\hat{\alpha}_o \hat{\varepsilon} - c\hat{\vec{\alpha}} \hat{\vec{p}}\right) = 0, \quad (2.7')$$

$$\left(\hat{\alpha}_o \hat{\varepsilon} + c\hat{\vec{\alpha}} \hat{\vec{p}}\right)\Psi = 0, \quad (2.7'')$$

which further we will conditionally name the linear semi-photon equations.

It is not difficult to show that only in the case when we choose the Dirac bispinors in the following form:

$$\Psi = \begin{pmatrix} E_x \\ E_z \\ iH_x \\ iH_z \end{pmatrix}, \quad \Psi^+ = \begin{pmatrix} E_x & E_z & -iH_x & -iH_z \end{pmatrix}, \qquad (2.8)$$

the equations (2.7) are the right Maxwell equations of the electromagnetic waves: retarded and advanced. Actually using (2.8) and putting in (2.7) we obtain:

$$\begin{cases} \dfrac{1}{c}\dfrac{\partial E_x}{\partial t} - \dfrac{\partial H_z}{\partial y} = 0 \\[4pt] \dfrac{1}{c}\dfrac{\partial H_z}{\partial t} - \dfrac{\partial E_x}{\partial y} = 0 \\[4pt] \dfrac{1}{c}\dfrac{\partial E_z}{\partial t} + \dfrac{\partial H_x}{\partial y} = 0 \\[4pt] \dfrac{1}{c}\dfrac{\partial H_x}{\partial t} + \dfrac{\partial E_z}{\partial y} = 0 \end{cases}, (2.9') \quad \begin{cases} \dfrac{1}{c}\dfrac{\partial E_x}{\partial t} + \dfrac{\partial H_z}{\partial y} = 0 \\[4pt] \dfrac{1}{c}\dfrac{\partial H_z}{\partial t} + \dfrac{\partial E_x}{\partial y} = 0 \\[4pt] \dfrac{1}{c}\dfrac{\partial E_z}{\partial t} - \dfrac{\partial H_x}{\partial y} = 0 \\[4pt] \dfrac{1}{c}\dfrac{\partial H_x}{\partial t} - \dfrac{\partial E_z}{\partial y} = 0 \end{cases}, \quad (2.9'')$$

(for waves of any other direction the same results can be obtained by the cyclic transposition of the indexes and by the canonical transformation of matrices and wave functions [2]).

In following we will conditionally name the equations (2.7)-(2.9) "the Maxwell-Dirac wave equations".

## 3.0. Twirl transformation of electromagnetic wave

The transformation of the linear wave to the curvilinear can be generally described by the following expression:

$$\hat{R}\Psi \to \psi, \qquad (3.1)$$

where $\hat{R}$ is the operator of some rotation transformation of linear EM wave; the $\Psi$ is the wave function, defined by matrix (2.8), which satisfies the equations (2.5) and (2.7), and $\psi$ is a yet unknown wave function:

$$\psi = \begin{pmatrix} E'_x \\ E'_z \\ iH'_x \\ iH'_z \end{pmatrix}, \qquad (3.2)$$

which has appeared after non-linear transformation, where $(E'_x, E'_z, H'_x, H'_z)$ are electromagnetic fields after transformation. Thus, our goal is from the equations (2.7) to find the equations, which satisfied the function (3.2).

As it is known, the description of vector transition from linear to curvilinear trajectory (briefly – "twirl transformation") is fully described by using of the differential geometry [3]. Note that mathematically this transition is equivalent to the vector transition from flat space to the curvilinear space, which is described by Riemann geometry.

(In connection to this let us remind that the Pauli matrices as well as the photon matrices are the space rotation operators – 2-D and 3-D correspondingly).

## 3.2. The twirl transformation description in electromagnetic form

Consider the equations (2.9). Let the plane-polarized wave, which has the field vectors $(E_x, H_z)$, be twirled with some radius $r_p$ in the plane $(X', O', Y')$ of a fixed co-ordinate system $(X', Y', Z', O')$ so that $E_x$ is parallel to the plane $(X', O', Y')$ and $H_z$ is perpendicular to it (figs 1 and 2).

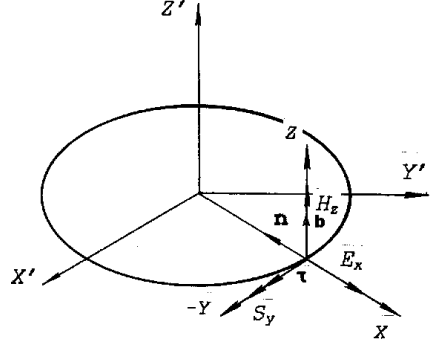

Fig. 2

According to Maxwell [1,4] the displacement current in the equation (2.9) is defined by the equation:

$$j_{dis} = \frac{1}{4\pi} \frac{\partial \vec{E}}{\partial t}, \qquad (3.3)$$

The above electrical field vector $\vec{E}$, which moves along the curvilinear trajectory (let it have direction from the center), can be written in the form:

$$\vec{E} = -E \cdot \vec{n}, \qquad (3.4)$$

where $E = |\vec{E}|$, and $\vec{n}$ is the normal unit-vector of the curve (having direction to the center). The derivative of $\vec{E}$ can be represented as:

$$\frac{\partial \vec{E}}{\partial t} = -\frac{\partial E}{\partial t} \vec{n} - E \frac{\partial \vec{n}}{\partial t}, \qquad (3.5)$$

Here the first term has the same direction as $\vec{E}$. The existence of the second term shows that the additional displacement current appears at the twirling of the wave. It is not difficult to show that it has direction, tangential to the ring:

$$\frac{\partial \vec{n}}{\partial t} = -v_p \kappa \vec{\tau}, \qquad (3.6)$$

where $\vec{\tau}$ is the tangential unit-vector, $v_p \equiv c$ is the electromagnetic wave velocity, $\kappa = \frac{1}{r_p}$ is the curvature of the trajectory and $r_p$ is some curvature radius. Thus, the displacement current of the plane wave, moving along the ring, can be written in the form:

$$\vec{j}_{dis} = -\frac{1}{4\pi} \frac{\partial E}{\partial t} \vec{n} + \frac{1}{4\pi} \omega_p E \cdot \vec{\tau}, \qquad (3.7)$$

where $\omega_p = \dfrac{m_p c^2}{\hbar} = \dfrac{v_p}{r_p} \equiv c\kappa$ we name the curvature angular velocity, $\varepsilon_p = m_p c^2$ is photon energy, $m_p$ is some mass, corresponding to the energy $\varepsilon_p$, $\vec{j}_n = \dfrac{1}{4\pi}\dfrac{\partial E}{\partial t}\vec{n}$ and $\vec{j}_\tau = \dfrac{\omega_p}{4\pi}E\cdot\vec{\tau}$ are the normal and tangent components of the current of the twirled electromagnetic wave, correspondingly.

### 3.1. The twirl transformation description in quantum form

Consider the Maxwell-Dirac wave equations (2.7) with wave function (2.8).

The generalization of the Dirac equation on the curvilinear (Riemann) geometry is connected with the parallel transport of the spinor in the curvilinear space [5,6,7]. We use below the most important results of this theory.

For the generalization of the Maxwell-Dirac equation on the Riemann geometry it is necessary to replace the usual derivative $\partial_\mu \equiv \partial/\partial x_\mu$ (where $x_\mu$ are the co-ordinates in the 4-space) with the covariant derivative:

$$D_\mu = \partial_\mu + \Gamma_\mu, \qquad (3.8)$$

where $\mu = 0, 1, 2, 3$ are the summing indices and $\Gamma_\mu$ is the analogue of Christoffel's symbols in the case of the spinor theory, called Ricci connection coefficients. In the theory it shown that $\hat{\alpha}_\mu \Gamma_\mu = \hat{\alpha}_i p_i + i\hat{\alpha}_0 p_0$, where $p_i$ and $p_0$ are the real values. It is not difficult to see that the tangent current $j_\tau$ corresponds to the Ricci connection coefficients (symbols) $\Gamma_\mu$.

When a spinor moves along the straight line, all the $\Gamma_\mu = 0$, and we have a usual derivative. But if a spinor moves along the curvilinear trajectory, not all the $\Gamma_\mu$ are equal to zero and a supplementary term appears. Typically, the last one is not the derivative, but it is equal to the product of the spinor itself with some coefficient $\Gamma_\mu$. Since, according to the general theory [8], the increment in spinor $\Gamma_\mu$ has the form and the dimension of the energy-momentum 4-vector, it is logical to identify $\Gamma_\mu$ with 4-vector of energy-momentum of the photon electromagnetic field:

$$\Gamma_\mu = \{\varepsilon_p, c\vec{p}_p\}, \qquad (3.9)$$

where $\varepsilon_p$ and $p_p$ is the photon energy and momentum (not the operators). In other words we have:

$$\hat{\alpha}_\mu \Gamma_\mu = \hat{\alpha}_0 \varepsilon_p + \vec{\hat{\alpha}}\,\vec{p}_p, \qquad (3.10)$$

Taking into account that according to energy conservation law $\hat{\alpha}_0 \varepsilon_p + \vec{\hat{\alpha}}\,\vec{p}_p = \pm\hat{\beta}\,m_p c^2$ it is not difficult to see that the supplementary term contains a twirled wave mass.

## 4.0. The equations of twirled electromagnetic wave

### 4.1. Klein-Gordon-like equation of twirling photon

As it is follows from previous sections due to the curvilinear motion of the electromagnetic wave, some additional terms $K$, corresponding to the tangent components of the displacement current, will appear in the equation (2.6), so that from (2.6) we have:

$$\left(\hat{\alpha}_o \hat{\varepsilon} - c\hat{\vec{\alpha}} \cdot \hat{\vec{p}} - K\right)\left(\hat{\alpha}_o \hat{\varepsilon} + c\hat{\vec{\alpha}} \cdot \hat{\vec{p}} + K\right)\psi = 0, \qquad (4.1)$$

where $K = \hat{\beta}\, m_p c^2$. Thus, in the case of the curvilinear motion of the electromagnetic field (photon) instead of the equation (2.6) we obtain the Klein-Gordon-like equation with mass [9]:

$$\left(\hat{\varepsilon}^2 - c^2 \hat{\vec{p}}^2 - m_p^2 c^4\right)\psi = 0, \qquad (4.2)$$

As we see the $\psi$-function, which appears after electromagnetic wave twirling and satisfies the equation (4.2), is not identical to the $\Psi$-function before twirling, which is the classical linear electromagnetic wave field and satisfies the equation (2.7).

As it is known in quantum physics the Klein-Gordon equation is considered as the scalar field equation. But obviously *the Klein-Gordon equation* (4.2), *whose wave function is 4-matrix with electromagnetic field components, cannot have the sense of the scalar field equation.* Actually, let us analyze the objects, which this equation describes.

From the Maxwell equations follows, that each of the components $E_x, E_y, E_z, H_x, H_y, H_z$ of vectors of an electromagnetic field $\vec{E}, \vec{H}$ submits to same form of the scalar wave equations. In the case of the linear waves all field components are independent. By study of one of the $\vec{E}, \vec{H}$ vectors components only, we can consider the vector field as scalar. But in case of CWED, when a tangential current appears, we cannot proceed to the scalar theory, since the components of a vector $\vec{E}$, as it follows from the condition $\vec{\nabla} \cdot \vec{E} = \dfrac{4\pi}{c} \vec{c}^{\,0} \cdot \vec{j}$, are not independent functions (here $\vec{c}^{\,o}$ is the unit vector of wave velocity).

Therefore, although the Klein - Gordon equation for scalar wave function describes a massive particle with spin zero (spinless boson), the equations (4.2) concerning electromagnetic wave functions (3.2), which appears after curvilinear transformation, represents the equation of the vector particle with rest mass $m_p$ and with unit spin (i.e. the particle similar to the intermediate boson), which we can name "heavy" photon.

Now we will analyse the particularities of the function $\psi = \{E'_x, E'_z, H'_x, H'_z\}$, appearing after twirl transformation, and the differences between electromagnetic fields $\{E_x, E_z, H_x, H_z\}$ of the $\Psi$ - function and electromagnetic fields $\{E'_x, E'_z, H'_x, H'_z\}$ of $\psi$-function.

### 4.2. The Dirac-like equation of the twirled electromagnetic wave

Taking into account the previous section results from (4.2) we can obtain the equations for the $\psi$-function, i.e. the twirled semi-photon equations:

$$[(\hat{\alpha}_o \hat{\varepsilon} + c\hat{\vec{\alpha}}\ \hat{\vec{p}}) + \hat{\beta}\ m_p c^2]\psi = 0, \qquad (4.3')$$

$$\psi^+[(\hat{\alpha}_o \hat{\varepsilon} - c\hat{\vec{\alpha}}\ \hat{\vec{p}}) - \hat{\beta}\ m_p c^2] = 0, \qquad (4.3'')$$

which is alike the Dirac electron-positron equations [9], but instead of electron mass, it contains the twirled photon mass $m_p$.

We can say that transition from equation (4.2) to the equation (4.3) is the twirled photon breaking into two charged particles, which must have opposite charge signs for the satisfaction of the charge conservation law. In the case of electron-positron pair production it must be $m_p = 2m_e$, where $m_e$ is the electron mass. Then from (4.3) we obtain:

$$[(\hat{\alpha}_o \hat{\varepsilon} + c\hat{\vec{\alpha}}\ \hat{\vec{p}}) + 2\hat{\beta}\ m_e c^2]\psi = 0, \qquad (4.4')$$

$$\psi^+[(\hat{\alpha}_o \hat{\varepsilon} - c\hat{\vec{\alpha}}\ \hat{\vec{p}}) - 2\hat{\beta}\ m_e c^2] = 0, \qquad (4.4'')$$

The question arises of why the equations (4.4) contain double masses relatively to the Dirac charge lepton equations? Let us analyze this problem.

Obviously after the twirled photon breaking, i.e. after the chargeless twirled photon is divided into two charged semi-photon, the plus and minus charged particles acquire the electric fields, and each particle begins to move in the field of another. In order to become independent (i.e. free) they must be drawn away the one from the other (fig.3)

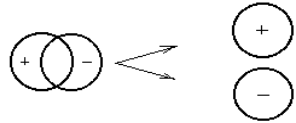

Fig.3

Therefore, the equations, which arise after the twirled photon equation division cannot be the free electron-positron equations, but the electron and positron equations with the external field.

Obviously, for the particle remotion from one another, the energy for the electric field creation must be expended. In fact, being the particles combined, the system doesn't have any field (fig. 3). At very small distance they create the dipole field. And at a distance, much more than the particle radius, the electron and positron acquire the full electric fields. As it is known [1,4], the potential $V_P$ of two point plus and minus charges in the point $P$ is defined as:

$$V_P = \frac{e}{4\pi}\left(\frac{1}{r} - \frac{1}{r + d\cos\theta}\right), \qquad (4.5)$$

where $\pm e$ is the charges, $d$ is the distance between the charges, and $\theta$ is the angle between axes and radius-vector of plus particle.

From this formula we can obtain the limit cases for two free particles:

$$\lim_{d \to \infty} V_P = \frac{1}{4\pi}\frac{e}{r}, \qquad (4.6)$$

During the breaking process the particle charges appear. For the particle, removed to infinity, the work against the attractive forces needed to be fulfilled:

$$\varepsilon_{rel} = \oint eV_P dv = \frac{1}{4\pi}\oint \frac{e^2}{r}dv, \qquad (4.7)$$

Obviously, this work is defined by the external particles field, so that the release energy is the field production energy and in the same time it is the annihilation energy. Therefore, due to energy conservation law the energy value for each particle must be equal $\varepsilon_{rel} = m_e c^2$.

Thus, the equations (4.3) we can write as electron equations with external field in the form:

$$\left[\left(\hat{\alpha}_0 \hat{\varepsilon} + c\hat{\vec{\alpha}}\ \hat{\vec{p}}\right) + \hat{\beta}\ m_e c^2 + \hat{\beta}\ m_e c^2\right]\psi = 0, \tag{4.8'}$$

$$\psi^+\left[\left(\hat{\alpha}_o \hat{\varepsilon} - c\hat{\vec{\alpha}}\ \hat{\vec{p}}\right) - \hat{\beta}\ m_e c^2 - \hat{\beta}\ m_e c^2\right] = 0, \tag{4.8''}$$

Using the linear equation of the energy conservation, we can write:

$$\pm \hat{\beta}\ m_e c^2 = -\varepsilon_{ex} - c\hat{\vec{\alpha}}\ \vec{p}_{ex} = -e\varphi_{ex} - e\hat{\vec{\alpha}}\ \vec{A}_{ex}, \tag{4.9}$$

where "*ex*" means "external". Putting (4.9) in (4.8) we obtain the Dirac equation with external field:

$$\left[\hat{\alpha}_0(\hat{\varepsilon} \mp \varepsilon_{ex}) + c\hat{\vec{\alpha}} \cdot \left(\hat{\vec{p}} \mp \vec{p}_{ex}\right) + \hat{\beta}\ m_e c^2\right]\psi = 0, \tag{4.10}$$

which at $d \to \infty$ give the Dirac free electron-positron equations:

$$\left[\left(\hat{\alpha}_o \hat{\varepsilon} + c\hat{\vec{\alpha}}\ \hat{\vec{p}}\right) + \hat{\beta}\ m_e c^2\right]\psi = 0, \tag{4.11'}$$

$$\psi^+\left[\left(\hat{\alpha}_o \hat{\varepsilon} - c\hat{\vec{\alpha}}\ \hat{\vec{p}}\right) - \hat{\beta}\ m_e c^2\right] = 0, \tag{4.11''}$$

From above follows some interesting consequences:
1. before breaking the twirled photon isn't an absolutely neutral particle, but a dipole; therefore, it must have the dipole momentum.
2. the formula (2.15) shows that in CWED the mass isn't equivalent to the energy, but to the 4-vector of the energy-momentum; from this follows that In CWED the energy has the kinetic origin.
3. in CWED the energy decomposition for free EM electron equation corresponds:

$$\pm \hat{\beta}\ m_e c^2 = -\varepsilon_{in} - c\hat{\vec{\alpha}}\ \vec{p}_{in} = -e\varphi_{in} - e\hat{\vec{\alpha}}\ \vec{A}_{in}, \tag{4.12}$$

where "*in*" means "internal". In other words the values $(\varepsilon_{in}, \vec{p}_{in})$ describe the inner field, and the values $(\varepsilon_{ex}, \vec{p}_{ex})$ the external field of electron. When we consider the electron from great distance, the field $(\varepsilon_{in}, \vec{p}_{in})$ works as the mass, and the term $(\varepsilon_{ex}, \vec{p}_{ex})$ describes the external electromagnetic field (and we have linear Dirac-like equations). Inside the electron the term $(\varepsilon_{in}, \vec{p}_{in})$ is needed for the detailed description of the inner field of an electron and carry to non-linear Dirac-like equations, as we will show below in section 8.

Obviously the equations (4.11) we can rewrite in form:

$$\frac{\partial \psi}{\partial t} - c\hat{\vec{\alpha}}\ \vec{\nabla}\psi - i\hat{\beta}\frac{c}{r_C}\psi = 0, \tag{4.13'}$$

$$\frac{\partial \psi}{\partial t} + c\hat{\vec{\alpha}}\ \vec{\nabla}\psi + i\hat{\beta}\frac{c}{r_C}\psi = 0, \tag{4.13''}$$

where $r_C = \dfrac{\hbar}{mc}$ is the Compton wavelength of the electron.

Using (3.2), from (4.11) or (4.13) we obtain electromagnetic form of these equations:

$$\begin{cases} \dfrac{1}{c}\dfrac{\partial E'_x}{\partial t} - \dfrac{\partial H'_z}{\partial y} = -ij^e_x \\ \dfrac{1}{c}\dfrac{\partial H'_z}{\partial t} - \dfrac{\partial E'_x}{\partial y} = ij^m_z \\ \dfrac{1}{c}\dfrac{\partial E'_z}{\partial t} + \dfrac{\partial H'_x}{\partial y} = -ij^e_z \\ \dfrac{1}{c}\dfrac{\partial H'_x}{\partial t} + \dfrac{\partial E'_z}{\partial y} = ij^m_x \end{cases} \quad (4.14') \qquad \begin{cases} \dfrac{1}{c}\dfrac{\partial E'_x}{\partial t} + \dfrac{\partial H'_z}{\partial y} = -ij^e_x \\ \dfrac{1}{c}\dfrac{\partial H'_z}{\partial t} + \dfrac{\partial E'_x}{\partial y} = ij^m_z \\ \dfrac{1}{c}\dfrac{\partial E'_z}{\partial t} - \dfrac{\partial H'_x}{\partial y} = -ij^e_z \\ \dfrac{1}{c}\dfrac{\partial H'_x}{\partial t} - \dfrac{\partial E'_z}{\partial y} = ij^m_x \end{cases} \quad (4.14'')$$

where

$$\begin{aligned} \vec{j}^e &= i\dfrac{\omega}{4\pi}\vec{E}' = i\dfrac{c}{4\pi}\dfrac{1}{r_C}\vec{E}' \\ \vec{j}^m &= i\dfrac{\omega}{4\pi}\vec{H}' = i\dfrac{c}{4\pi}\dfrac{1}{r_C}\vec{H}' \end{aligned} \qquad (4.15)$$

are the imaginary currents, in which $\omega = \dfrac{mc^2}{\hbar}$.

As we see the equations (4.14) are Maxwell equations with imaginary currents, which differ by the directions. (As it is known the existence of the magnetic current $\vec{j}^m$ doesn't contradict to the quantum theory; see the Dirac theory of the magnetic monopole [10]). In our case (see figs. 2 and 3) the z-components of the currents are equal to zero.

Thus, the equations (4.11) are Maxwell equations with imaginary tangential currents and simultaneously they are the Dirac-like equation of EM-particle with non-zero mass and charge, and half spin.

## 5.0. Electromagnetic form of the free electron equation solution

In accordance with the above results the electromagnetic form of the solution of the Dirac-like free EM-electron equation must be a twirled electromagnetic wave. Let us show that this supposition is actually correct.

From the above point of view for the $y$-direction photon two solutions must exist:
1) for the wave, twirled around the $OZ$-axis

$${}^{oz}\psi = \begin{pmatrix} E'_x \\ 0 \\ 0 \\ iH'_z \end{pmatrix} = \begin{pmatrix} \psi_1 \\ 0 \\ 0 \\ \psi_4 \end{pmatrix}, \qquad (5.1)$$

and 2) for the wave, twirled around the $OX$-axis

$$^{ox}\psi = \begin{pmatrix} 0 \\ E'_z \\ iH'_x \\ 0 \end{pmatrix} = \begin{pmatrix} 0 \\ \psi_2 \\ \psi_3 \\ 0 \end{pmatrix}, \quad (5.2)$$

Let us compare (5.1) and (5.2) with the Dirac electron theory solutions [9].
It is known [9] that the solution of the Dirac free electron equation (2.1) has the form of the plane wave:

$$\psi_j = B_j \exp\left(-\frac{i}{\hbar}(\varepsilon t - \vec{p}\vec{r})\right) \quad (5.3)$$

where $j = 1, 2, 3, 4$; $B_j = b_j e^{i\phi}$; the amplitudes $b_j$ are the numbers and $\phi$ is the initial wave phase. The functions (5.3) are the eigenfunctions of the energy-momentum operators, where $\varepsilon$ and $\vec{p}$ are the energy-momentum eigenvalues. Here for each $\vec{p}$, the energy $\varepsilon$ has either positive or negative values according to the energy conservation law equation $\varepsilon_\pm = \pm\sqrt{c^2\vec{p}^2 + m^2c^4}$.

For $\varepsilon_+$ we have two linear-independent set of four orthogonal normalizing amplitudes:

$$1)\ B_1 = -\frac{cp_z}{\varepsilon_+ + mc^2},\ B_2 = -\frac{c(p_x + ip_y)}{\varepsilon_+ + mc^2},\ B_3 = 1,\ B_4 = 0, \quad (5.4)$$

$$2)\ B_1 = -\frac{c(p_x - ip_y)}{\varepsilon_+ + mc^2},\ B_2 = \frac{cp_z}{\varepsilon_+ + mc^2},\ B_3 = 0,\ B_4 = 1, \quad (5.5)$$

and for $\varepsilon_-$:

$$3)\ B_1 = 1,\ B_2 = 0,\ B_3 = \frac{cp_z}{-\varepsilon_- + mc^2},\ B_4 = \frac{c(p_x + ip_y)}{-\varepsilon_- + mc^2}, \quad (5.6)$$

$$4)\ B_1 = 0,\ B_2 = 1,\ B_3 = \frac{c(p_x - ip_y)}{-\varepsilon_- + mc^2},\ B_4 = -\frac{cp_z}{-\varepsilon_- + mc^2}, \quad (5.7)$$

Let's analyze these solutions.
**1)** The existing of two linear independent solutions corresponds with two independent orientations of the electromagnetic wave vectors and gives the unique logic explanation for this fact.
**2)** Since $\psi = \psi(y)$, we have $p_x = p_z = 0$, $p_y = mc$ and for the field vectors we obtain: from (4.4) and (4.5) for "positive" energy

$$B_+^{(1)} = \begin{pmatrix} 0 \\ b_2 \\ b_3 \\ 0 \end{pmatrix} \cdot e^{i\phi},\ B_+^{(2)} = \begin{pmatrix} b_1 \\ 0 \\ 0 \\ b_4 \end{pmatrix} \cdot e^{i\phi}, \quad (5.8)$$

and from (4.6) and (4.7) for "negative" energy:

$$B_-^{(1)} = \begin{pmatrix} b_1 \\ 0 \\ 0 \\ b_4 \end{pmatrix} \cdot e^{i\phi}, \quad B_-^{(2)} = \begin{pmatrix} 0 \\ b_2 \\ b_3 \\ 0 \end{pmatrix} \cdot e^{i\phi}, \qquad (5.9)$$

which exactly correspond to (5.1) and (5.2).

**3)** Calculate the correlations between the components of the field vectors. Putting $\phi = \dfrac{\pi}{2}$ for $\varepsilon_+ = mc^2$ and $\varepsilon_- = -mc^2$ we obtain correspondingly:

$$B_+^{(1)} = \begin{pmatrix} 0 \\ \frac{1}{2} \\ i \cdot 1 \\ 0 \end{pmatrix}, \quad B_+^{(2)} = \begin{pmatrix} -\frac{1}{2} \\ 0 \\ 0 \\ i \cdot 1 \end{pmatrix}, \qquad (5.10)$$

$$B_-^{(1)} = \begin{pmatrix} i \cdot 1 \\ 0 \\ 0 \\ -\frac{1}{2} \end{pmatrix}, \quad B_-^{(2)} = \begin{pmatrix} 0 \\ i \cdot 1 \\ \frac{1}{2} \\ 0 \end{pmatrix}, \qquad (5.11)$$

Obviously the imaginary unit in these solutions indicates that the field vectors $\vec{E}$ and $\vec{H}$ are mutualy orthogonal.

Also we see that the electric field amplitude is two times less, than the magnetic field amplitude. This fact shows that the electromagnetic fields, which correspond to the Dirac-like equation solution, are different contrary to fields of the linear wave of the Maxwell theory, where $\vec{E} = \vec{H}$. (It can be show that this result provides the EM-electron stability).

**4)** It is easy to show that the electromagnetic form of the solution of the Dirac equation is the standing wave. Really in case of the twirled wave we have $\vec{p} \perp \vec{r}$ and therefore $\vec{p} \cdot \vec{r} = 0$; then instead (4.3) we obtain:

$$\psi_j = b_j \exp\left(-\frac{i}{\hbar}\varepsilon\, t\right), \qquad (5.12)$$

**5)** According with the Euler formula $e^{i\varphi} = \cos\varphi + i\sin\varphi$ the solution of the Dirac-like equation (5.12) describes a circle, as it corresponds to our theory.

## 6.0. Specification of wave function of non-linear EM Dirac-like equation

Let us try now to specify a connection between the electromagnetic wave, which satisfies the Dirac-like equation, and the initial photon wave.

As is known [11,12], the fields of a photon are vectors and will be transformed according to elements of group (O3). The spinor fields of the Dirac equation will be transformed as elements of group (SU2). As it is shown by L.H. Ryder [11], two spinor transformations correspond to one transformation of a vector. For this reason the spinors are also named "semi-vectors" or " tensors of half rank " [7,8].

From above following that the twirling and breaking of the photon-like waves corresponds to transition from usual linear Maxwell equation to the Maxwell equation for the curvilinear wave with an imaginary tangential current (i.e. to the EM Dirac-like equation). Obviously, the transformation properties of electromagnetic fields at this transition change. As wave functions of the Dirac equation (i.e. spinors) submit to transformations of group (SU2), the semi-photon fields will submit to the same transformations.

Thus, in the case of half spin particles creation, the non-linear (twirl) transformation is followed by spontaneous breaking of the twirled electromagnetic wave:

$$\left(\hat{R}\Psi\right) \to \psi \begin{array}{c} \to \psi \\ \to \psi^+ \end{array}, \tag{6.1}$$

Now a question arises about which particularities the semi-photon fields have and what do they differ for from the photon fields (i.e. an electromagnetic wave).

The answer to the above question follows from the analysis of transformation properties of the twirled semi-photon wave function. Taking into account that we have the same mathematical equations both for the CWED Dirac equation and the Dirac lepton equation, we can affirm that these transformation features coincide with the same features of the spinor [11,13].

The spinor transformation has the form:

$$\psi' = U\psi, \tag{6.2}$$

where the operator of transformation is entered as follows:

$$U(\vec{n}\theta) = \cos\frac{1}{2}\theta - i\vec{n}\cdot\vec{\sigma}'\sin\frac{1}{2}\theta, \tag{6.3}$$

where $\vec{n}$ is the unit vector of an axis, $\theta$ is a rotation angle around this axis and $\vec{\sigma}' = (\sigma_x', \sigma_y', \sigma_z')$ is the spin vector.

The rotation matrix (6.3) possesses a remarkable property. If the rotation occurs on the angle $\theta = 2\pi$ around any axis (therefore occurs the returning to the initial system of reference) we find, that $U = -1$, instead of $U = 1$ as it was possible to expect. Differently, the state vector of system with spin half, in usual three-dimensional space has ambiguity and passes to itself only after turn to the angle $4\pi$.

This result can be explained only if we suppose that the *EM lepton-like particle of the above EM Dirac-like equation is the twirled half-period of a twirled photon particle, and therefore needs to be rotated twice to return to the initial state*.

Taking into account the results of previous section it is appropriate to note here that the solution of the EM Dirac-like equation in the electromagnetic form we can name "electromagnetic spinor". In other words the electromagnetic spinor is the semi-period of twirling electromagnetic wave.

Thus the transformation of the linear electromagnetic wave into curvilinear wave and its breaking produces the electromagnetic spinors.

## 7.0. Illustration of the twirl transformation

The process of CWED electron-like and positron-like particles production by the twirled photon breaking into two twirled half-periods must be illustrated in following way (fig. 4):

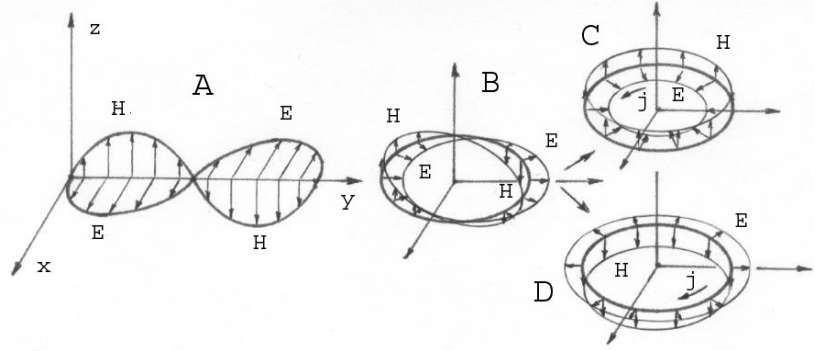

Fig. 4

As we see the "daughter" twirled semi-photons C and D are completely anti-symmetric and can't be transformed one to each other by any transformation of co-ordinates (if it is not to be accompanied by the changing of field direction).

It is clear that here (fig. 4) the parts *C* and *D* contain the currents of opposite directions. Thus we can suppose, that the cause of a twirled photon breaking is the mutual repulsion of oppositely directed currents.

It is not difficult to understand that the above breaking process corresponds to the quantum process of the particle-antiparticle pair production. Actually the transformation (disintegration) of the massless quantum of an electromagnetic wave $\gamma$ into two massive particles in the presence of strong EM nuclear field $Ze$, i.e. the electron-positron $e^-, e^+$ pair production process

$$\gamma + Ze \to e^+ + e^- + Ze, \qquad (7.1)$$

can be considered in parallel to the process of EM-string twirling and breaking. In the case of such identification we can calculate some features of this EM particles, which are some particularities of CWED.

It is interesting that *both the twirled semi-photon and the twirled photon radii must be the same*. This fact follows from the angular momentum conservation low. Really, the twirled photon angular momentum is equal to:

$$\sigma_p = p_p \cdot r_p = 2m_e c \cdot \frac{\hbar}{2m_e c} = 1\hbar, \qquad (7.2)$$

In accordance with the momentum conservation low:

$$\sigma_s^+ + \sigma_s^- = \sigma_p, \qquad (7.3)$$

where $\sigma_s^+, \sigma_s^-$ are the spins of the plus and minus semi-photons (i.e. of the EM electron and EM positron fields; the index "s" characterizes further the semi-photon parameters). Then we obtain:

$$\sigma_s = \frac{1}{2}\sigma_h = \frac{1}{2}\hbar, \qquad (7.4)$$

Since

$$\sigma_s = p_s \cdot r_s, \qquad (7.5)$$

where $r_s$ is the twirled semi-photon (electron) radius, and $p_s = m_e c$ is the inner semi-photon (electron) momentum, we have

$$r_s = \frac{\sigma_s}{p_s} = \frac{1}{2}\frac{\hbar}{m_e c} = \frac{\hbar}{2m_e c} = r_p, \qquad (7.6)$$

Thus, *the circle size of the twirled photon doesn't change after breaking*.
As it is not difficult calculate, *the angular velocity (angular frequency) also doesn't change*:

$$\omega_s = \frac{c}{r_s} = \frac{2m_e c^2}{\hbar} = \omega_p.$$

The breaking of the twirled photon makes it possible to outline the solution of the some known theoretical problems:

*The difference between positive and negative charges*: this difference follows from the difference of the field and tangent current direction of the twirled semi-photons after pair production (by condition that the Pauli - principle is true).

*Zitterbewegung*. The results obtained by E. Schroedinger in his well-known articles about the relativistic electron [14] are the most important confirmation for the electron structure model. He showed, that electron has a special inner motion "Zitterbewegung", which has frequency $\omega_Z = \frac{2m_e c^2}{\hbar}$, amplitude $r_Z = \frac{\hbar}{2m_e c}$, and velocity of light $\upsilon = c$. The attempts to explain this motion had not given results. But if the electron is the twirled semi-photon, then we receive the simple explanation of Schroedinger's analysis.

*Infinite problems*: in the non-linear theory the charge and mass infinite problems don't exist.

*The electric charge appearance*: is the consequence of the occurrence of a tangential displacement current of Maxwell at a photon twirling and breaking.

*The spin of EM particles* arises owing to the twirling of the field of an electromagnetic wave.

Additionally from above follows the results, which fully explain the many experimental results of physics:

*The origin of the charge conservation low*: since in nature there are the same numbers of the photon half periods, the sum of the particles charge is equal zero.

*The difference between bosons and fermions* is well explained: the bosons contain the even number and the fermions contain the odd number of the twirled semi-photons.

*Masses of EM particles*. In CWED the masses of the particles arise as the "stopped" electromagnetic energy of a twirled photon and semi-photon. Therefore only the linear photon rest mass is equal to zero; all other particles should have a non-zero rest mass.

*The neutrality of the Universe*: the quantities of "positive" and "negative" half-periods are equal in the Universe.

*The existence of EM particles and antiparticles* corresponds to the antisymmetry of the twirled semi-photons.

*The spontaneous breakdown of symmetry of EM wave field and the occurrence of the EM particle mass* take place at the moment of its twisting and breaking into two half-periods at presence of the nuclear field. As Higgs field here work the electromagnetic field of the nucleus.

*The EM twirled wave - particle duality* ceases be a riddle in the theory of the twirled photons: EM particles really represent simultaneously both waves and particles.

## 8.0. Non-linear Dirac-like equation and Lagrangian

### 8.1. The nonlinear Dirac-like equation

Obviously the curvi-linearity means non-linearity. The stability of twirled photon is possible only by the photons parts self-action. From this follows that the CWED Dirac-like equation must be the non-linear electromagnetic field equation. Let us find its form.

Using (4.12) from (4.11) we will obtain the following non-linear equation:

$$[\hat{\alpha}_0(\hat{\varepsilon} - \varepsilon_{in}) + c\hat{\vec{\alpha}} \cdot (\hat{\vec{p}} - \vec{p}_{in})]\psi = 0, \quad (8.1)$$

where the inner energy and momentum can be expressed by the inner energy density and momentum density (or by Poynting vector) $U$, $\vec{g}$, $\vec{S}$ of EM wave correspondingly:

$$\varepsilon_{in} = \int_0^\tau U \, d\tau = \frac{1}{8\pi}\int_0^\tau (\vec{E}^2 + \vec{H}^2) \, d\tau, \quad (8.2)$$

$$\vec{p}_{in} = \int_0^\tau \vec{g} \, d\tau = \frac{1}{c^2}\int_0^\tau \vec{S} \, d\tau = \frac{1}{4\pi}\int_0^\tau [\vec{E} \times \vec{H}] \, d\tau, \quad (8.3)$$

putting the upper limit $\tau$ to be variable.

Actually, substituting of the expression (8.2) and (8.3) to the Dirac-like electron equation, we obtain the non-linear integral-differential equation, which is, as we suppose, the searched form of the nonlinear equation, which describes the EM-electron in both electromagnetic and concurrent quantum forms. To clear out the mathematical particularities of the equation (8.1) we should find its approximate quantum form.

It is easy to prove that for the EM wave the quantum forms of $U$ and $\vec{S}$ are:

$$U = \frac{1}{8\pi} \psi^+ \hat{\alpha}_0 \psi, \quad (8.4)$$

$$\vec{S} = -\frac{c}{8\pi} \psi^+ \hat{\vec{\alpha}} \psi = c^2 \vec{g}, \quad (8.5)$$

Taking into account that the free electron Dirac equation solution is the plane wave:

$$\psi = \psi_0 \exp[i(\omega t - ky)], \quad (8.6)$$

we can write (8.2) and (8.3) in the next aproximate form:

$$\varepsilon_p = U \, \Delta\tau = \frac{\Delta\tau}{8\pi} \psi^+ \hat{\alpha}_0 \psi, \quad (8.7)$$

$$\vec{p}_p = \vec{g} \, \Delta\tau = \frac{1}{c^2}\vec{S} \, \Delta\tau = -\frac{\Delta\tau}{8\pi c} \psi^+ \hat{\vec{\alpha}} \psi, \quad (8.8)$$

where $\Delta\tau$ is the volume, which contain the main part of the twirled semi-photon energy.

Then the approximate form of the equation (8.3) is following:

$$\frac{\partial \psi}{\partial t} - c\hat{\vec{\alpha}}\vec{\nabla}\psi + i\frac{\Delta\tau}{8\pi c}(\psi^+ \hat{\alpha}_0 \psi - \hat{\vec{\alpha}}\psi^+ \hat{\vec{\alpha}}\psi)\psi = 0, \quad (8.9)$$

It is not difficult to see that the equation (8.9) is the nonlinear equation of the same type as following non-linear Heisenberg equation:

$$\gamma_\mu \frac{\partial \psi}{\partial x_\mu} + \frac{1}{2}l[\gamma_\mu \psi(\overline{\psi}\gamma_\mu \psi) + \gamma_\mu \gamma_5 \psi(\overline{\psi}\gamma_\mu \gamma_5 \psi)] = 0, \quad (8.10)$$

if instead of $\alpha$-set matrices we use $\gamma$-set matrices (here $l$ is some positive constant). The equation (8.10) was investigated by Heisenberg et. al. [15,16] and played for a while the role of the unitary field theory equation. Contrary to the last one, the equation (8.9) is obtained in a logical and correct way and the self-action constant $l$ appeared in (8.9) automatically. As it is known in the framework of this non-linear unitary field theory some substantial achievements were made [15,16].

## 8.2. The Lagrangian density of the nonlinear EM Dirac-like equation

The Lagrangian density of the linear Dirac equation in quantum form is [9]:

$$L_D = \psi^+ \left( \hat{\varepsilon} + c\hat{\vec{\alpha}} \cdot \hat{\vec{p}} + \hat{\beta} m_e c^2 \right) \psi, \tag{8.11}$$

or in the electromagnetic form:

$$L_D = \frac{\partial U}{\partial t} + div \, \vec{S} - i \frac{\omega}{8\pi} \left( \vec{E}^2 - \vec{H}^2 \right), \tag{8.12}$$

(Note that in the case of the variation procedure we must distinguish the complex conjugate field vectors $\vec{E}^*, \vec{H}^*$ and $\vec{E}, \vec{H}$).

The Lagrangian density of nonlinear equation is not difficult to obtain from the Lagrangian density of the linear Dirac equation [9] using the method by which we found the nonlinear equation. By substituting (5.1) we obtain:

$$L_N = \psi^+ \left( \hat{\varepsilon} - c\hat{\vec{\alpha}} \cdot \hat{\vec{p}} \right) \psi + \psi^+ \left( \varepsilon_{in} - c\hat{\vec{\alpha}} \cdot \vec{p}_{in} \right) \psi, \tag{8.13}$$

We suppose that the expression (8.13) represents the common form of the Lagrangian density of the nonlinear twirled electromagnetic wave equation. In order to compare (8.13) with the known results of classical and quantum physics let us find the approximate electromagnetic and quantum forms of this equation.

Using (8.7) and (8.8) we can represent (8.11) in the quantum form:

$$L_N = i\hbar \left[ \frac{\partial}{\partial t} \left[ \frac{1}{2} \left( \psi^+ \psi \right) \right] - c \, div \left( \psi^+ \hat{\vec{\alpha}} \psi \right) \right] + \frac{\Delta \tau}{8\pi} \left[ \left( \psi^+ \psi \right)^2 - \left( \psi^+ \hat{\vec{\alpha}} \psi \right)^2 \right], \tag{8.14}$$

By the normalizing $\psi$-function by the expression $L'_N = \frac{1}{8\pi \, mc^2} L_N$ and transforming (8.13) in the electrodynamics form, using equations (8.4) and (8.5), we will obtain from (8.14) the following approximate electromagnetic form:

$$L'_N = i \frac{\hbar}{2m_e} \left( \frac{1}{c^2} \frac{\partial U}{\partial t} + div \, \vec{g} \right) + \frac{\Delta \tau}{m_e c^2} \left( U^2 - c^2 \vec{g}^2 \right), \tag{8.15}$$

It is not difficult to transform the second term, using the known electrodynamics transformation:

$$(8\pi)^2 \left( U^2 - c^2 \vec{g}^2 \right) = \left( \vec{E}^2 + \vec{H}^2 \right)^2 - 4 \left( \vec{E} \times \vec{H} \right)^2 = \left( \vec{E}^2 - \vec{H}^2 \right)^2 + 4 \left( \vec{E} \cdot \vec{H} \right)^2, \tag{8.16}$$

Thus, taking into account that $L_D = 0$ and using (8.12) and (8.16), we obtain from (8.15) the following expression:

$$L'_N = \frac{1}{8\pi} \left( \vec{E}^2 - \vec{H}^2 \right) + \frac{\Delta \tau}{(8\pi)^2 mc^2} \left[ \left( \vec{E}^2 - \vec{H}^2 \right)^2 + 4 \left( \vec{E} \cdot \vec{H} \right)^2 \right], \tag{8.17}$$

As we see, the approximate form of the Lagrangian density of the nonlinear equation of the twirled electromagnetic wave contains only the invariants of the Maxwell theory and is similar to the known Lagrangian density of the photon-photon interaction [17].

Let us now analyze the quantum form of the Lagrangian density (8.17). The equation (5.12) can be written in the form:

$$L_Q = \psi^+ \hat{\alpha}_\mu \partial_\mu \psi \ + \frac{\Delta \tau}{8\pi}\left[ \left(\psi^+ \hat{\alpha}_0 \psi\right)^2 - \left(\psi^+ \hat{\vec{\alpha}}\ \psi\right)^2 \right], \tag{8.18}$$

It is not difficult to see that the electrodynamics correlation (8.16) in quantum form has the known form of the Fierz identity [18]:

$$\left(\psi^+ \hat{\alpha}_0 \psi\right)^2 - \left(\psi^+ \hat{\vec{\alpha}}\ \psi\right)^2 = \left(\psi^+ \hat{\alpha}_4 \psi\right)^2 + \left(\psi^+ \hat{\alpha}_5 \psi\right)^2, \tag{8.19}$$

Using (8.19) from (8.18) we obtain:

$$L_Q = \psi^+ \hat{\alpha}_\mu \partial_\mu \psi \ + \frac{\Delta \tau}{8\pi}\left[ \left(\psi^+ \hat{\alpha}_4 \psi\right)^2 - \left(\psi^+ \hat{\alpha}_5 \psi\right)^2 \right], \tag{8.20}$$

The Lagrangian density (8.20) coincides with the Nambu and Jona-Lasinio Lagrangian density [19], which is the Lagrangian density of the relativistic superconductivity theory. As it is known this Lagrangian density is used for the solution of the problem of the elementary particles mass appearance by the mechanism of the vacuum symmetry spontaneous breakdown (it corresponds also to the Cooper's pair production in the superconductivity theory).

**Conclusion**

We have showed that there is non-linear EM field equation, which mathematically is fully concurrent to the quantum electrodynamics equation.